\documentstyle[prb,twocolumn,aps,tighten, floats,epsfig]{revtex}
\begin{document}
\twocolumn[\hsize\textwidth\columnwidth\hsize\csname @twocolumnfalse\endcsname

\title{Dynamic modulation of electron correlation by intramolecular modes in charge transfer compounds}
\maketitle
\vskip 0.25 truein
\centerline{M. Meneghetti}
\centerline{Department of Physical Chemistry, University of Padova, 2, Via Loredan,I-35131 Padova, Italy}
\vskip 0.25 truein]
\noindent {\bf Electron-phonon and electron-electron interactions are in competition in determining the properties of molecular charge transfer conductors and superconductors. The direct influence of phonons on the electron-electron interaction was not before considered and in the present work 
the coupling of intramolecular modes to electron--electron interaction  ($U$-vib interaction) is investigated.
The effect of this coupling on the frequency of the normal modes of a dimer model is obtained and it is  shown that frequency shifts of the Raman active modes are directly related to this coupling. The results are used to obtain the values of the $U$-vib coupling constants of intramolecular modes of a representative molecule of charge transfer conductors, like tetramethyltetratiafulvalene. Consequences of this coupling on the electron pairing are also suggested.}
\vskip 1pc
The interesting electronic and optical behaviors of  organic $\pi$-conjugated systems derive from their low energy electronic properties. One recalls, for example, the superconducting properties of some charge transfer molecular crystals or the luminescent properties of some $\pi$-conjugated polymers used for producing light emitting diodes.\cite{ICSM96} The interaction of electrons with phonons has  to be considered, in particular for these small band systems, as important as  the Coulomb interaction between the $\pi$ electrons. In fact some properties, like a low symmetry ground state due to a Peierls or a spin-Peierls transition,\cite{Pytte,Hirsh} derive from the electron-phonon interaction, but other properties like the charge transfer excitations are dominated by the Coulomb interaction.\cite{Mazumdar}  On the other hand  it is not available a satisfactory model for the supercondunting properties of  these systems based on a well defined interaction.  

Hubbard or extended Hubbard models have been extensively used for describing the electronic properties of $\pi$-conjugated systems since they allow to describe the electron-electron interaction beyond the mean field approximation.\cite{Baeriswyl} Electron-phonon interaction has been included in such Hamiltonians\cite{Rice1,Bozio1,Meneghetti1} on the basis of the Holstein model\cite{Holstein} which considers the dependence of the site energies on the {\it intra}-molecular normal modes and of  the SSH model,\cite{SSH} derived from the Peierls model,\cite{Peierls} which takes into account the dependence of the transfer integrals on the {\it inter}-molecular modes. It is the aim of this paper to find evidences of the possible dependence of the intrasite Coulomb interaction of the Hubbard model on the {\it intra}-molecular vibrations ($U$-vib interaction). This interaction can be of direct importance for understanding the dynamic of electrons in low dimensional systems like organic charge transfer superconductors since it can favor the pairing of electrons if the modulation of the Coulomb interaction is comparable to that of the interaction itself.

The simplest model for studying the $U$-vib interaction is a half-filled charge transfer dimer with two identical molecules, namely two electrons on two sites. Since it is well known that one can easily obtain charge transfer dimers of radical molecules in solution, one has also the possibility of directly verifying the results of the model. We will consider in particular the cation radicals of tetramethyltetrathiafulvalene (TMTTF$^{\dot+}$), a molecule which is present in many charge transfer crystrals and in one superconducting compound.\cite{Balicas}

The Hubbard Hamiltonian for the dimer is
\begin{eqnarray}
H=&&\epsilon \sum_{i=1,2} n_i 
+ t \sum_\sigma(a^\dagger_{1,\sigma} a^{}_{2,\sigma}  +a^\dagger_{2,\sigma} a^{}_{1,\sigma})\nonumber\\
&&+ U \sum_{i=1,2}\sum_\sigma n_{i,\sigma}n_{i,-\sigma} 
\label{Hub}
\end{eqnarray}
where $\epsilon$ is the site energy and can be taken as zero, $t$ is the hopping integrals and $U$ is the intramolecular Coulomb interaction. 

The interaction with the intramolecular normal modes of vibration is introduced by expanding, to first order, $\epsilon$ and $U$ on the basis of the normal modes. Because of the symmetry of the dimer, the following coordinates, for each $Q_\alpha$ intramolecular normal mode, are used 
$
Q^S_\alpha=2^{-1/2}(Q_{\alpha,1}+ Q_{\alpha,2})
$ 
and 
$
Q^A_\alpha=2^{-1/2}(Q_{\alpha,1}-Q_{\alpha,2})
$.
On this basis the following two terms are added to  Eq. \ref{Hub}:
\begin{equation}
\sum_\alpha Q^A_\alpha [{g_\alpha \over \sqrt{2}}(n_2 - n_1) + {g^U_\alpha \over \sqrt{2}}(n_{1,\uparrow}n_{1,\downarrow}- n_{2,\uparrow}n_{2,\downarrow})]
\end{equation}
and
\begin{equation}
\sum_\alpha Q^S_\alpha {g^U_\alpha \over \sqrt{2}}(n_{1,\uparrow}n_{1,\downarrow}+ n_{2,\uparrow}n_{2,\downarrow})
\end{equation}
where $g_\alpha=(\partial \epsilon/\partial Q_\alpha)_\circ$ is the coupling constant based on the Holstein model and $g^U_\alpha=(\partial U/\partial Q_\alpha)_\circ$ is that for the $U$-vib interaction. 

The contribution coming from the Holstein mechanism involves only  the $Q^A_\alpha$ coordinates which, by symmetry, can be observed in the infrared spectra, whereas the $U$-vib interaction affects also the  $Q^S_\alpha$ coordinates which are Raman active. Previously, attention was directed in particular to the infrared spectra where large and important effects have been observed.\cite{Bozio2} However, one can see that it is in the Raman spectra that one may find the sign of the $U$-vib interaction and it is in particular to these spectra of the charge transfer dimer that we direct our attention.

\begin{figure}[htb]
\centerline{\psfig{figure=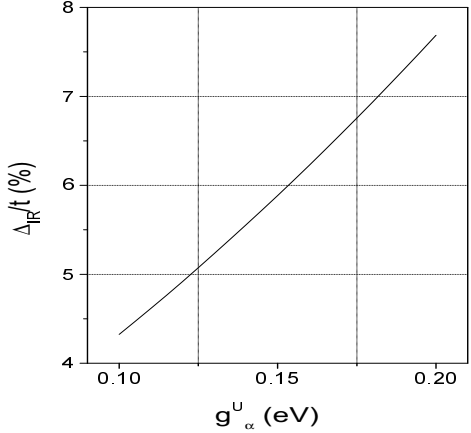,width=3.0 in,height=3.0in}}
\caption{Frequency shift of the infrared active $Q^A_\alpha$ mode as a function of $g^U_\alpha$ with $U$=1.0 eV, $t$=0.2 eV and $g_\alpha$=0.1 eV.}
\label{IRSh}
\end{figure}

The evaluation of the electron-vibration interaction strength can be obtained from the frequency shifts of the vibrational modes observed in the infrared and Raman spectra. The case in which only the Holstein mechanism is present has already been considered.\cite{Rice2,Meneghetti2} In this case the vibrational shift can be calculated  for the infrared active $Q^A_\alpha$ modes. The only difference  one should consider in the present case for these coordinates is that the coupling constant for the $\alpha$-th mode change from $g_\alpha$ to $(g_\alpha+g^U_\alpha/2)$. The result,\cite{Meneghetti2} can now be rewritten in a more convenient form as:
\begin{equation}
\Delta_{IR}= \hbar\omega_\alpha -\hbar\omega_{IR}\simeq {4  |\Delta_{ST}| (g_\alpha + g^U_\alpha/2)^2 \over
             \hbar \omega_{CT}  (\hbar \omega_{CT} +|\Delta_{ST}|)}
\label{deltaIR}
\end{equation}
where $\hbar\omega_\alpha$ and $\hbar\omega_{IR}$ are the unperturbed and perturbed energy, respectively, of the $\alpha$-th mode, $\hbar\omega_{CT} =U/2+(U^2/4 +4 t^2)^{1/2}$ is the energy of the charge transfer excitation of the dimer and $\Delta_{ST}=U/2-(U^2/4 +4 t^2)^{1/2}$ is the energy separation between the singlet ground state and the triplet states. One finds that a coupled $Q^A_\alpha$ mode shifts to lower frequencies.

\begin{figure}[htb]
\centerline{\psfig{figure=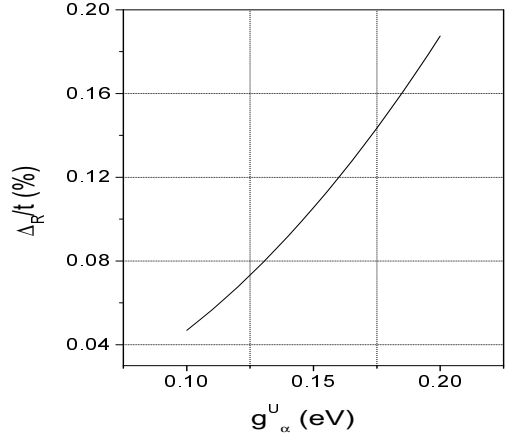,width=3.0 in,height=3.0in}}
\caption{As for Fig. \protect\ref{IRSh} but for the Raman active $Q^S_\alpha$ mode.}
\label{RSh}
\end{figure}

The frequencies of the $Q^S_\alpha$ modes, which can be observed in the Raman spectra, were previously considered as the frequencies of the unperturbed modes since these modes were not involved in the electron-vibration interaction. Now, if $U$-vib interaction is present, these modes are no more unperturbed and following the derivation of Eq. \ref{deltaIR},\cite{Meneghetti2} one finds that their vibrational energy shift are:
\begin{equation}
\Delta_R=\hbar\omega_\alpha-\hbar\omega_{R}\simeq {|\Delta_{ST}|^2 (g^U_\alpha)^2 \over(\hbar \omega_{CT} +|\Delta_{ST}|)^3}.
\label{deltaR}
\end{equation}

The reference, unperturbed state, which defines $\hbar\omega_\alpha$, must now be the isolated radical molecule, namely the situation in which the charge transfer excitation is not present.

Before looking at the experimental spectra it is important to obtain some indications about the vibrational frequency shifts predicted by the model.
Figures \ref{IRSh} and \ref{RSh} show the calculated frequency shifts for the infrared ($Q^A_\alpha$) and Raman ($Q^S_\alpha$) modes, respectively, {\it vs.} the coupling constant $g^U_\alpha$ for values of the other parameters characteristic of a charge transfer system: $U$=1.0 eV, $t$=0.2 eV and $g_\alpha$=0.1 eV.    

One notes that the shifts for the infrared mode are  more than one order of magnitude larger than those calculated for the Raman mode. The calculation shows that one can observe vibrational shift as large as 100 cm$^{-1}$ in infrared, as it was found many times,\cite{Bozio2} but that in the Raman spectrum one can predict  vibrational shifts of the order of a few wavenumbers. This could be the reason why there have not been observations of frequency shifts of the Raman active modes for segregated systems like the dimer we are considering.

The small frequency shift one calculates for the Raman mode suggests  care in obtaining the experimental data.  In particular it is important that both the unperturbed and the perturbed frequency can be recorded  using the same sample and in the same experiment so that to minimize possible influence of the experimental conditions. The opportunity of doing such an experiment is given by the equilibrium between the isolated molecules and the charge transfer dimers,  which is present in solution of radical molecules like TMTTF$^{\dot+}$. From the absorptions in the visible and near-infrared spectral regions it is easy to identify the presence of the isolated molecules and of the charge transfer dimers.\cite{Torrance} 

\begin{figure}
\centerline{\psfig{figure=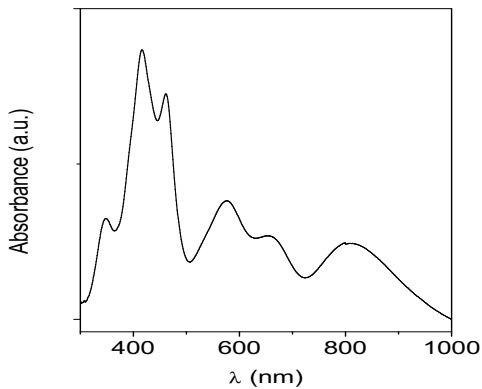,width=3.0 in,height=3.0in}}
\caption{Electronic spectrum of TMTTF$^{\dot+}$ in sulfolane. Both the transitions of the monomer (670 and 460 nm) and of the dimer (805 (CT), 575, 417 nm) are observed.}
\label{Vis-NIR}
\end{figure}

Figure \ref{Vis-NIR} shows the electronic spectra of a sulfolane solution of TMTTF$^{\dot+}$  radical cation in which the electronic transitions of the monomer and of the dimer are easily recognizable. The absorption at longer wavelength (805 nm) is the charge transfer excitation of the dimers. The electronic excitations at 670 nm and at 460 nm are those of the monomer and they shift to shorter wavelength (575 nm and 417 nm) when the dimer is formed.\cite{Torrance} 
Since the absorption bands of the monomer and of the dimer are found at sufficiently different energies, one can obtain the Resonance Raman spectra of the two species using the appropriate excitation laser frequency. One finds that the Krypton laser lines at 568 and at 647 nm are in resonance with the electronic transitions of the dimer and of the monomer, respectively. 

The same equilibrium between the monomer and the dimer  can be found using dimethylsulfoxide (DMSO) as solvent. We have also obtained samples in which the dimers of TMTTF$^{\dot+}$ are embedded in a polymer like PMMA. In these cases, however, the equilibrium is shifted almost completely toward the dimers and it is not possible to obtain both the spectrum of the monomer and of the dimer using the same material.  

The Raman spectrum of TMTTF$^{\dot+}$ shows two strong bands at 530 and at 1430 cm$^{-1}$, and another one at 1057 cm$^{-1}$. The bands have been assigned as $a_g$ $\nu_{10}$, $ a_g$ $\nu_{4}$ and to an internal vibration of the methyl groups, respectively.\cite{Meneghetti3} ${a_g}$ $\nu_{10}$ was described as a  displacement of the internal C-S bonds whereas ${a_g}$ $\nu_{4}$ was related to the stretching of the internal C=C bond. From the infrared data of dimerized TMTTF$^{\dot+}$ systems it was concluded\cite{Meneghetti3} that these two normal modes were strongly coupled to the charge transfer excitation since two strong new bands, which were not expected for the isolated molecule, can be observed in the infrared spectra and assigned to the infrared activation of these two normal modes.

The Raman spectra of the solutions have been recorded by using a Jobin-Yvon multichannel spectrometer (S3000) equipped with a CCD camera (Astromed) cooled at liquid nitrogen. The spectra of the solutions were obtained at room temperature with a  rotating cell for liquids to avoid the decomposition of TMTTF$^{\dot+}$ due to a poor dissipation of the energy of the laser beam. The solutions were prepared by using (TMTTF)ClO$_4$ and were flushed with nitrogen to eliminate the presence of oxygen. We used about 40 mW and a beam waist estimated to be 200$\mu$m. It was verified that there was no decomposition of the solutions by recording the UV-Visible spectra after the Raman measurements. It was found that the reproducibility of the frequencies of the spectra was better than 1 cm$^{-1}$. 

\begin{figure}
\centerline{\psfig{figure=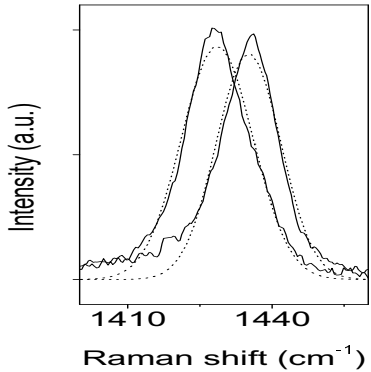,width=3.0 in,height=2.5in}}
\caption{Raman spectrum of the  $ a_g$ $\nu_{4}$ mode of TMTTF$^{\dot+}$. The peak at higher frequency has been recorded with the 647 nm  exciting laser line of a Krypton laser in resonance with a monomer electronic transition, whereas that at lower frequency has been recorded with the line at 568 nm, in resonance with a dimer transition. The dotted lines are fitting of the experimental data with a single gaussian function. }
\label{Raman}
\end{figure}

The spectra of the sulfolane solutions showed two different behaviors for the $a_g$ $\nu_{10}$ at about 530 cm$^{-1}$ and for the $ a_g$ $\nu_{4}$ at about 1430 cm$^{-1}$.  The first band at 530 cm$^{-1}$ showed a small positive shift of less then one wavenumbers when the exciting line was changed from 647 to 568 nm, namely from the resonance with the monomer to that of the dimer. On the contrary, the second band, reported in Figure \ref{Raman}, was found at 1435.1 cm$^{-1}$  when the resonance was with the monomer electronic transition and at 1428.6 cm$^{-1}$ when the resonance was with the dimer one. Therefore  in this case one finds a frequency shifts of -6.5 cm$^{-1}$. The  fitting of the bands with gaussian functions showed that a single band was present in every case without a significant variation in the widths. 

The same situation was found when the Resonance Raman spectra of the DMSO solutions were recorded. In this case the frequencies of the modes were  different due to the different environment (for the monomer we observed 524.1 and 1430.3 cm$^{-1}$ for $a_g$ $\nu_{10}$ and $ a_g$ $\nu_{4}$ respectively). However, we found that the frequency shifts of the two modes were similar to those of the previous situation with a small positive shift of 1.2 cm$^{-1}$ for $a_g$ $\nu_{10}$  and a negative shift  of -5.0 cm$^{-1}$ for  $ a_g$ $\nu_{4}$. 

The evidence was, therefore, that a significant  shift is observed, when the dimer is formed, only for the mode related to the stretching of the central C=C bond of the molecule. The observed frequency shift is in agreement with the theory prediction of a small negative shift for a normal modes coupled with the charge transfer excitation. 

However, to rule out the possibility that the formation of the dimer may induce some variation, not related to the charge transfer excitation, which could influence the vibrational normal modes, we have recorded the Raman spectra of the solution and of the solid of the neutral  molecule. The  crystal structure  of the neutral molecule is not available but from these spectra one could reasonably obtain some indications about the difference between an almost free molecule in solution and one which is in close contact with other molecules. The $a_g$ $\nu_{4}$ mode has been observed, for the neutral molecule, at 1538 cm$^{-1}$ for the solid at about 20 K.\cite{Meneghetti3} One finds that, at room temperature, its frequency is 1532.5 cm$^{-1}$  for the crystalline material and  less than one wavenumber higher for the solution in carbon sulfide in which the neutral molecule dissolves sufficiently. This can be taken as an indication that the aggregation alone has not a significant influence on the frequency of the normal mode related to the C=C central bond. 

By using  Eq. \ref{deltaR} one can estimate the $g^U_\alpha$ coupling constant for $a_g$ $\nu_{4}$. The frequency of the charge transfer excitation is readly found in the near infrared spectrum where it can be located, below  all the intramolecular electronic transitions, at 805 nm. $\Delta_{ST}$, the energy separation between the ground state and the  triplet states, can be measured in an EPR experiment in which the intensity of the  lines associated to the Boltzman population of the triplet states, is measured against temperature. For the dimers of TMTTF$^{\dot +}$ these magnetic measurements are available.\cite{Meneghetti4} The dimers were embedded in PMMA where one can verify that almost only the dimeric species is present. It was found that $\Delta_{ST}=0.25 $ eV. 

These data allow one to find $|g^U_4|=256 \pm 17$ meV, considering the minimum and maximum observed Raman shift. With this value one can also find the value for the Holstein coupling $g_4$ by using Eq. \ref{deltaIR} for the frequency shift of the infrared normal mode. In this case it is also important the relative sign of the two coupling constants $g_4$ and $g^U_4$ which, however, cannot be determined on the basis of the above equations. Therefore one takes into account that the sign of the two coupling constant may be different. 

The frequency of the $Q^A_4$ infrared mode was already found to be  1361 cm$^{-1}$ for a solution of dimers of TMTTF$^{\dot+}$ in DMSO.\cite{Meneghetti3} Since the frequency of the unperturbed mode observed in the Raman spectrum of the DMSO solution is 1430.3 cm$^{-1}$, one finds that the  frequency shift for $ Q^A_{4}$ is 69.3  cm$^{-1}$. This shift is consistent with the predictions reported above on the basis of the dimer model and allow us to calculate $g_4$. 

One finds that if the two coupling constant have the same sign 
$|g_4|=32\pm 8$ meV whereas in the case of different signs  $|g_4|=285\pm 8$ meV. The reported errors are only indicative of the large and small values of $g^U_4$. Previously, the value of $|g_4|$ was reported to be 120 meV.\cite{Pedron}  Therefore, we find that the new estimate for $g_4$ are very different from the previous ones and this could be important for the dynamic of the electron transfer between the molecules. 

In any case one finds a large $g^U_\alpha$, namely
an important dynamic modulation of the correlation term of the Hubbard Hamiltonian ($U$) which can be evaluated, from the value of $\hbar\omega_{CT}$ and the value of $\Delta_{ST}$ to be 1.29 eV. This situation is very interesting since it can be important for the formation of local pairs of electrons and this opens the way for an explanation of the superconducting properties of organic charge transfer salts following the condensation of bipolarons.\cite{Ranninger}  One should recall, however, that such a superconducting ground state was never observed for a half-filled charge transfer compound and that this property was only found for quarter filled systems.\cite{Ishiguro} In this case the relevant Coulomb interaction is not $U$ but $V$, namely the interaction between electrons on nearest neighbor sites, since there is one electron, or hole, on every second site . Therefore one could suggest that in this case it is the modulation of $V$ which is important for a pairing mechanism ($V$-vib interaction), a modulation that can derive both from the {\it intra}- and the {\it inter}-molecular vibrations. Values of $V$, for charge transfer compounds, have been found to be of the order of the transfer integral\cite{Mazumdar,Meneghetti1} and one can speculate that the modulation could be very large, due to the coupling to both type of vibrations, determining a strong dynamical pairing of electrons on neighboring sites. It would be therefore very important to obtain values of this type of electron-vibration interaction.

In conclusion, we have obtained, for the first time, experimental evidences that in organic charge transfer compounds  there is a coupling between electronic correlation and intramolecular vibrations ($U$-vib interaction). In particular we have studied the Raman spectra of dimers of a representative molecule like TMTTF$^{\dot+}$ and found that its $ a_g$ $\nu_{4}$ normal mode strongly modulates ($|g^U_4|=256 \pm 17$ meV) the value of the intramolecular Coulomb interaction. This result shows that also the value of the coupling constants related to the Holstein mechanism ($g_\alpha$) have to be reconsidered. The value of $g^U_4$ seems to be too small to conclude that a pairing of electrons in organic superconductors follows from this interaction, but suggests that experimental data about the same mechanism but involving intersite Coulomb correlation ($V$-vib), have to be obtained.

I wish to thank Renato Bozio for valuable discussions and  Gabriele Marcolongo and Gilda Deluca for technical help.
Financial support by the Italian National Research Council also through its Progetto Finalizzato MSTA2 and the Ministry of the University and of Scientific and Technological Research are acknowledged.


\begin{references}
\bibitem[*]{byline}E-mail: M.Meneghetti@chfi.unipd.it
\bibitem{ICSM96} Proceedings of the International Conference on Science and Technology of Synthetic Metals, edited by Z. V. Vardeny, A. J. Epstein [Synth. Met. {\bf 84}-{\bf 86} (1997)].
\bibitem{Pytte} E. Pytte, Phys. Rev. B {\bf 10}, 4637 (1974).
\bibitem{Hirsh} H. J. Hirsh, in {\it Low dimensional conductors and superconductors}, edited by D. Jerome and L. R. Caron  (Nato ASI Series,  1986), Vol. 155, p. 71.
\bibitem{Mazumdar} F. B. Gallagher and S. Mazumdar, Phys. Rev. B {\bf 56}, 15025 (1997).
\bibitem{Baeriswyl} D. Baeriswyl, D. K. Campbell, and S. Mazumdar, in {\it Conjugated Conducting Polymers}, edited by H. Kiess (Springer Verlag, Heidelberg, 1992), p. 7.
\bibitem{Rice1} M. J. Rice, N. O. Lipari, and S. Str\"assler, Phys. Rev. Lett. {\bf 39}, 1359 (1977).
\bibitem{Bozio1}R. Bozio, M. Meneghetti, and C. Pecile, Phys. Rev. B {\bf 36}, 7795 (1987); D. Pedron, A. Speghini, V. Mulloni, and R. Bozio, J. Chem. Phys. {\bf 103}, 2795 (1995).
\bibitem{Meneghetti1} M. Meneghetti, Phys. Rev. B {\bf 44}, 8554 (1991).
\bibitem{Holstein} T. Holstein, Ann. Phys. {\bf 8}, 325 (1959).
\bibitem{SSH}A. J. Heeger, S. Kivelson, J. R. Shrieffer, and W.-P. Su, Rev. Mod. Phys. {\bf 60}, 781 (1988).
\bibitem{Peierls} R. E. Peierls, {\it Quantum Theory of solids}, Clarendon, Oxford, 1955, p. 108.
\bibitem{Balicas} L. Balicas, K. Behnia, W. Kang, P. Auban-Senzier, E. Canadell, D. Jerome, M. Ribault, and J-M. Fabre, Adv. Mat. {\bf 6} 792 (1994).
\bibitem{Bozio2} R. Bozio and C. Pecile, in {\it Spectroscopy of advanced materials}, edited by R. J. Clark and R. E. Hester (John Wiley \& Sons, 1991), p. 1.
\bibitem{Rice2} M. J. Rice, Solid State Commun. {\bf 31}, 93 (1979).
\bibitem{Meneghetti2} M. Meneghetti and C. Pecile, Phys. Rev. B {\bf 42}, 1605 (1990).
\bibitem{Torrance}J. B. Torrance, B. A. Scott, B. Welber, F. B. Kaufman, and P. E. Seiden, Phys. Rev. B {\bf 19} 730 (1979).
\bibitem{Meneghetti3} M. Meneghetti, R. Bozio, I. Zanon, C. Pecile, C. Ricotta, and M. Zanetti,  J. Chem. Phys. {\bf 80}, 6210 (1984).
\bibitem{Meneghetti4} M. Meneghetti, A. Toffoletti, and L. Pasimeni, Phys. Rev. B {\bf 54}, 16353 (1996).
\bibitem{Pedron} D. Pedron, R. Bozio, M.Meneghetti, and C. Pecile, Phys. Rev. B {\bf 49}, 10893 (1994). 
\bibitem{Ranninger}R. Micnas, J. Ranninger, S. Robaszkiewicz, Rev. Mod. Phys. {\bf 62}, 113 (1990).
\bibitem{Ishiguro} T. Ishiguro, K. Yamaji, {\it Organic superconductors} (Springer Verlag Berlin, 1990).


\end{references}
\end{document}